# Tailoring the atomic structure of graphene nanoribbons by STM lithography

Levente Tapasztó[1]*, Gergely Dobrik[1], Philippe Lambin[2], and László P Biró[1]

1. Research Institute for Technical Physics and Materials Science, H-1525 Budapest, Hungary,
2. Facultes Universitaire Notre Dame de la Paix, 61 Rue de Bruxelles, B-5000 Namur, Belgium

* email: tapaszto@mfa.kfki.hu

**The practical realization of nano-scale electronics faces two major challenges: the precise engineering of the building blocks and their assembly into functional circuits [1]. In spite of the exceptional electronic properties of carbon nanotubes [2] only basic demonstration-devices have been realized by time-consuming processes [3, 4, 5]. This is mainly due to the lack of selective growth and reliable assembly processes for nanotubes. However, graphene offers an attractive alternative. Here we report the patterning of graphene nanoribbons (GNRs) and bent junctions with nanometer precision, well-defined widths and predetermined crystallographic orientations allowing us to fully engineer their electronic structure using scanning tunneling microscope (STM) lithography. The atomic structure and electronic properties of the ribbons have been investigated by STM and tunneling spectroscopy measurements. Opening of confinement gaps up to 0.5 eV, allowing room temperature operation of GNR-based devices, is reported. This method avoids the difficulties of assembling nano-scale components and allows the realization of complete integrated circuits, operating as room temperature ballistic electronic devices[6,7].**

High electron mobility and long coherence length make graphene a subject of intense focus for nano-scale electronic applications even for the realization of room-temperature ballistic (dissipation-free) electronics[8]. However, a major setback in the development of graphene-based field effect transistors is the inability to electrostatically confine electrons in graphene, since a single layer of graphite remains metallic even at the charge neutrality point[9]. In order to overcome this problem, a way to open a gap in the electronic structure of graphene has to be found. A straightforward solution is to pattern the graphene sheet into narrow ribbons, which can be viewed as unrolled single-walled carbon nanotubes. Due to the quantum mechanical constriction of electronic wave functions in the direction perpendicular to the axis of the ribbon, a confinement-induced gap can open. Theoretical works predicted the strong dependence of the gap on width and crystallographic orientation of the ribbon[10,11]. This dependence offers us the opportunity to tailor the electronic structure of the graphene nanoribbons. However, it also imposes strict requirements for the fabrication of GNR-based devices, since in order to provide sufficient reproducibility, the accurate control of their structure with nanometer precision is imperative.

With standard e-beam lithographic methods, GNRs down to a few tens of nanometers in width could be realized recently[12, 13], which seems to be the limit of this technique[1]. However, GNRs of a few nanometers width are required in order to obtain energy gaps adequate for room temperature operation, since the gap scales inversely with ribbon width[10, 12]. It is obvious that this scale cannot be achieved by e-beam lithography. Moreover, electron lithographic methods have difficulties in controlling the crystallographic orientation of the ribbons.

Scanning probe microscopy techniques and in particular STM might offer the solution. STM combines the capability of atomic resolution imaging with the ability to locally modify the surface of the samples[14]. The modification of graphite surface by STM has its almost twenty-year old history[15]. However, previous works focused on studying the mechanism of surface modification by analyzing the pit formation in graphite. Although the potential for etching lines was also demonstrated [16] no further significant advance has been reported. In our work we combine the ability of surface modification of STM with atomic resolution imaging in order to engineer nanostructures with almost atomically precise structure and predetermined electronic properties.

Cutting of GNRs was done by applying a constant bias potential (significantly higher than the one used for imaging) and simultaneously moving the STM tip with constant velocity in order to etch the desired geometry fitted to the crystallographic structure known from previous atomic resolution STM imaging (see the Methods section for details).

Figure 1.a shows a 10 nm wide and 120 nm long graphene nanoribbon etched by STM lithography. By setting the optimal lithographic parameters (2.4 V bias potential and 2.0 nm/s tip velocity) we were able to cut GNRs with suitably regular edges, which constitutes a great advance regarding the reproducibility of GNR-based devices. Furthermore, more complex graphene nano-architectures can be tailored by STM lithography. As a basic demonstration, Fig. 1 b shows an 8 nm wide, 30 degree GNR-bent-junction connecting an armchair and a zigzag

ribbon, giving rise to a metal-semiconductor molecular junction[7]. After the patterning process, in situ atomic resolution STM images could be achieved revealing the atomic structure of the GNRs. The ribbon shown in Fig. 2 a is about 15 nm wide and its axis has a crystallographic orientation close to the zigzag direction. The atomic resolution image reveals an atomically flat and defect-free structure far from the edges, where $\sqrt{3}\times\sqrt{3}\,R30°$ type superstructure patterns can be observed. The origin of these patterns is the interference of the electrons scattered at the irregularities of the edges[17, 18] (Fig. 2 e). For electronic device applications these ribbons have to be deposited on an insulating substrate. However, for atomic resolution STM and Scanning Tunneling Spectroscopy (STS) investigation of GNRs the surface of highly oriented pyrolytic graphite (HOPG) provides an ideal substrate, since it allows the ribbon to remain flat without inducing additional defects or changing its topology[19,20,21]. The intrinsic characteristics of GNRs can be studied this way.

When imaging a 10 nm wide armchair GNR at low bias voltage (100 mV), oscillations in the electron density distribution parallel to the axis of the ribbon, reminding a Fabry-Perot electron resonator, were observed as shown in Fig. 3 a. The periodicity of the observed oscillations was about 0.4 nm, which clearly differs from the period of the underlying atomic structure (0.246 nm), and corresponds to the Fermi wavelength of electrons in graphene (Fig. 3 b). These findings are also supported by Fourier transformed images clearly indicating the presence of a lower frequency oscillation (red circles) as compared to the atomic periodicity (yellow circles) in Fig. 3 c. To correctly interpret the measurements we have performed theoretical modeling of STM images of GNRs based on a tight-binding $\pi$-electron Hamiltonian method[22]. The method was successfully used before in the interpretation of atomic resolution STM images of carbon nanotubes. Figure 4 shows the calculated constant current STM image of an 1.7 nm wide armchair GNR. Stripes of high electron density distribution running parallel to the axis of the ribbon are clearly visible. The period of these oscillations is about 0.37 nm close to the experimentally observed periodicity of 0.41 nm measured in Fig. 3. The small difference between calculated and measured values might occur due to the doping of GNR edges in air. Since at low bias voltages the STM measurements map the (square modulus of the) electronic wave function near the Fermi level[23], we attribute these oscillations to the quantum mechanical confinement of electrons across the ribbon. The presence of continuous interference patterns along the whole length of the ribbon in the experimental images is a spectacular evidence of phase coherent quantum billiard (a standing electron wave) in GNRs at room temperature (RT), evidencing their behavior as electronic waveguides even under ambient conditions. Both real and reciprocal space images (Fig. 3b and 3c respectively) reveal a single characteristic oscillation period, indicating the one dimensional (1D) nature of electronic structure of the narrow GNRs. Larger area and bias dependent STM images are provided as supplementary information in Figure S4 and S2 respectively.

In order to experimentally investigate the electronic band structure of GNRs, STS measurements have been preformed. Representative STS spectra taken on the 10 nm wide armchair GNR are shown in Fig. 3 d. The dI/dV quantity, which is proportional to the local electronic density of states, revealed the presence of van Hove singularities, a signature of a 1D electronic structure, as well-known for CNTs [24]. Comparison with LDOS of HOPG is given in supplementary information Fig. S3. Furthermore, the distance between the first pair of van Hove singularities gives the value of the energy gap[24]. For the 10 nm wide armchair GNR, a 0.18 eV gap value was found, which seems to suit well the theoretical rule for separation of the energy levels due to the geometrical constriction of wave functions[25] $E_g(W) = \pi \hbar v_0 / W \approx (2 eV \cdot nm)/W$, where $W$ is the width of the ribbon, and $v_0$ the Fermi velocity of electrons in graphene. Our results concerning the electronic wave-guide behavior of GNRs are also in accordance with those derived from transport measurements[25,26] at low temperatures and for much wider ribbons.

Another advantage of STM lithography beyond the precise engineering of GNR's structure is the potential for downscaling. We were able to etch GNRs down to 2.5 nm width corresponding to 10 carbon ring units along the width of the ribbon (Fig 5 a). The parameters used for lithography were 2.28 V bias potential and 1.0 nm/s tip velocity. The width and crystallographic orientation of GNR shown in Fig. 5 a correspond to an unrolled (10,0) zigzag SWCNT. The STS measurements performed on the 2.5 nm wide ribbon shown in Fig. 5 c reveal an energy gap of about 0.5 eV in excellent agreement with first principles theoretical calculations[11]. Tight binding calculations predict a gap of $1.2 eV \cdot nm/W$ which is also in good agreement with our results. Furthermore, the gap value of 0.5 eV is comparable with that of Ge (0.67 eV), allowing the RT operation of GNR-based electronic devices.

However, unlike CNTs, which are seamless graphitic structures rolled into a perfect cylinder, GNRs have edges. Since very narrow GNRs are needed to achieve the desired gap, the effect of edges can be critical. It was theoretically shown, that edge irregularities might induce electronic states within the gap region[27]. Our STM measurements also revealed oscillations in the electronic density of states of the 2.5 nm wide ribbon. These oscillations are less regular than the ones observed for wider ribbons, their orientation encloses a 30 degree angle with the ribbon axis, and the corresponding energy lies within the expected gap, which suggests that the imaged states are probably related to edge disorder.

Although SPM methods are usually characterized by a low throughput, the manufacturing of large scale integrated circuits of graphene by the SPM lithographic process presented above is not unrealistic, as "Millipede" type[28] scanning probe microscopes (SPM) are able to work in parallel with more than 1000 tips which can significantly increase the efficiency.

In summary, we have developed an STM lithography based technology that allows the engineering of graphene with true nanometer precision. We etched semiconducting ribbons from graphene with predetermined energy gap values up to 0.5 eV allowing their operation at room temperature. These GNR devices also show a phase coherent behavior even under ambient conditions. Furthermore, STM lithography offers us the opportunity for patterning more complicated architectures, even complete integrated circuits from networks of GNRs. One can imagine the feasibility of a variety of electronic devices based on our technique, which could open up new directions in the experimental realization of graphene-based electronics.

**Methods**

In order to pattern a single graphene sheet on the surface of a HOPG sample, we used a commercial DI Nanoscope III STM operating under ambient conditions. Pt-Ir tips proved to be the most suitable for both imaging and lithography. First, atomic resolution images were taken on the atomically flat graphene sheet. Then the sample was rotated in order to set the desired crystallographic orientation of the ribbon axis (edges). Afterwards, the graphene layer was cut by applying a constant bias potential (significantly higher than the one used for imaging) and simultaneously moving the STM tip with constant velocity in order to etch the desired geometry. Good results were obtained for positive sample biases. The microscopic mechanism of etching is not yet fully understood. Most likely the breaking of carbon-carbon bonds by field emitted electrons combined with the electron-transfer-enhanced oxidation of the graphene is responsible for the etching[29]. The whole lithographic process was controlled by a custom-written computer code enabling us to manage several parameters. The applied bias potential and the velocity of the tip during the patterning were found to be the critical parameters. The optimal parameters were slightly dependent on the microstructure of the mechanically etched tips but typically varied in the range of 2.2 – 2.6 V bias voltages and 1.0 – 5.0 nm / s tip velocities. Here we note that during the patterning we used tip velocities which were three orders of magnitude lower than those reported in Ref 16, and moved the tip only once along the lines to be etched, instead of scanning it several hundreds of times with high velocity. After optimizing the parameters of the lithographic process we were able to cut single layer (0.335 nm) deep, down to 1 nm wide, and several hundreds of nanometers long trenches with well-defined edges. Experimental evidence of patterning only a single graphene layer is shown in Fig. S1 of supplementary material. The STM lithographic patterning showed a good stability and reproducibility even under ambient conditions. After several etching processes the quality of the STM tip was still good enough to achieve atomic resolution images of the GNRs, enabling us to in situ investigate their atomic structure.

**Acknowledgements**

This work was supported in Hungary by OTKA Grant No. 67851 and OTKA-NKTH Grant K67793.

Correspondence and requests for materials should be addressed to L.T.

**Author contributions**

L.T. conceived the experiments: L.T. and G. D. performed the experiments: L.T., P.L. and L.P.B. analyzed the data, L.T. wrote the paper. All authors discussed the results and commented on the manuscript.

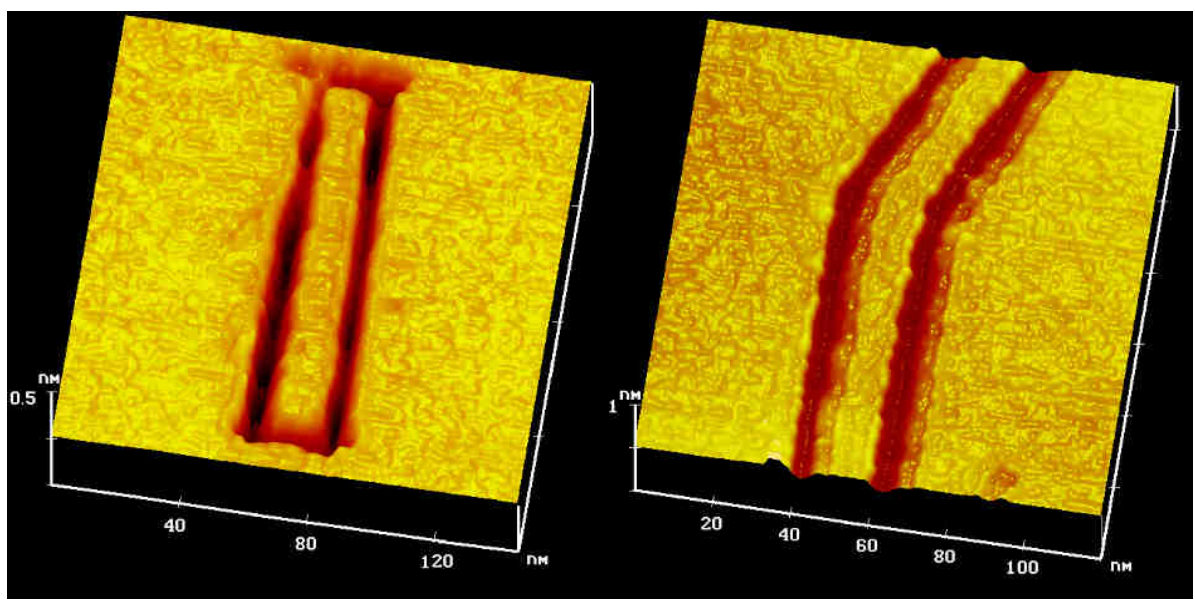

**Figure 1** Graphene nanostructures patterned by STM lithography. **a)** 3D STM image of a 10 nm wide and 120 nm long graphene nanoribbon. **b)** An 8 nm wide 30° GNR bent-junction connecting an armchair and a zigzag ribbon.

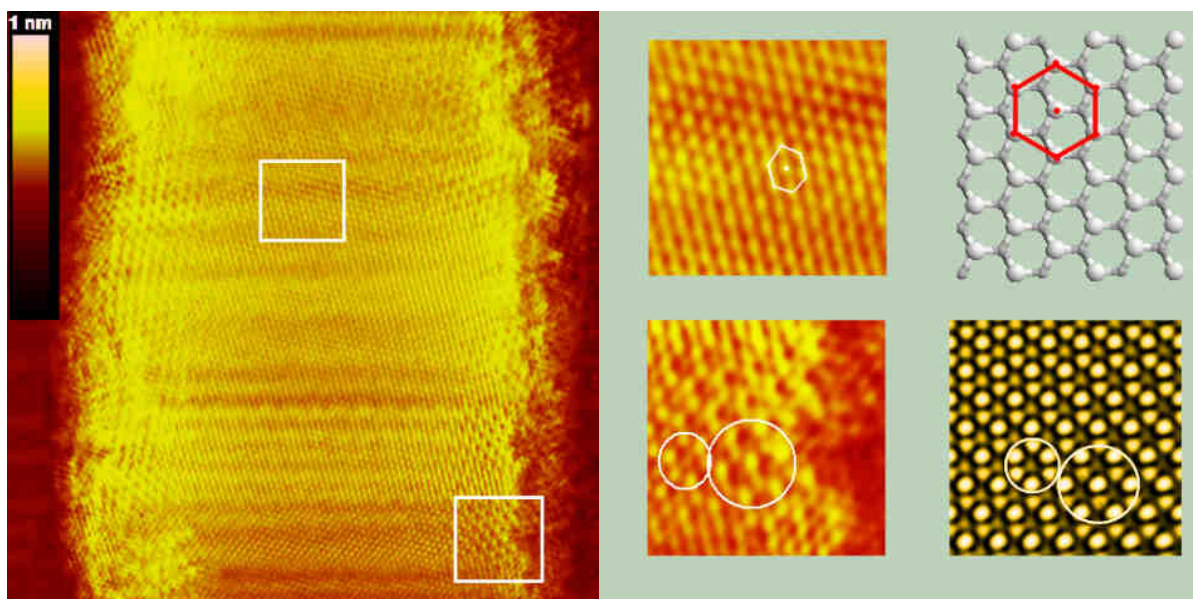

**Figure 2** Atomic structure of graphene nanoribbons. **a)** Atomic resolution STM image (20x20 nm$^2$ , 1 nA, 200 mV) of a 15 nm wide GNR displaying an atomically flat and defect-free structure. Magnified images of the: **b)** defect-free lattice taken at the center of the ribbon and **c)** position dependent superstructures near the edges. **d)** Identification of crystallographic orientation from the triangular lattice observed in atomic resolution STM images of HOPG supported GNR. **e)** Theoretical STM image of the superstructures at the edges of the ribbon.

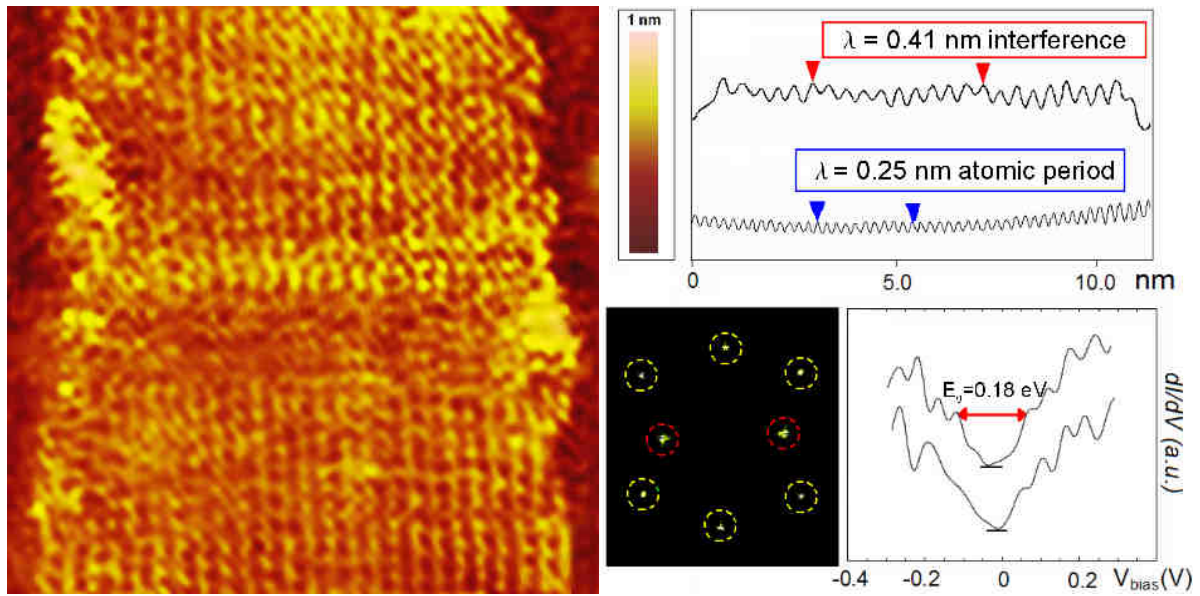

**Figure 3** Electronic structure of GNRs. **a)** Constant current STM image (12x12 nm$^2$, 1 nA, 100 mV) of a 10 nm wide armchair GNR displaying confinement induced interference patterns (stripes parallel to the axis of the ribbon). **b)** Average line-cuts revealing the period of the observed oscillation, which clearly differs from the periodicity of the underlying atomic structure. **c)** 2D Fourier transformation of the STM image. **d)** Representative STS spectra taken on the ribbon revealing an energy gap of 0.18 eV (zero DOS marked by horizontal lines).

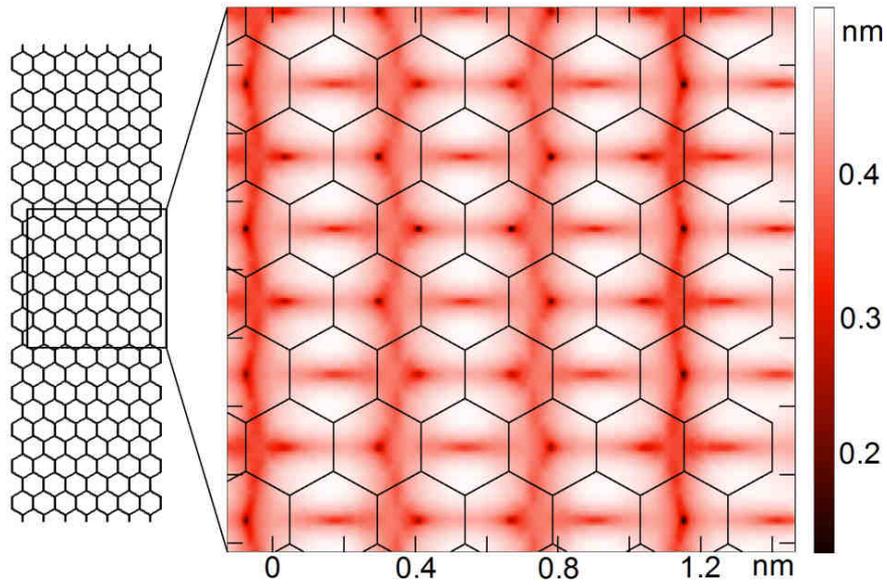

**Figure 4** Tight-binding computation of the STM image of GNRs. Topographical STM image at constant current calculated for a 1.7 nm wide armchair GNR at $V_t = 0.5$V tip potential and 0.5 nm tip-GNR distance.

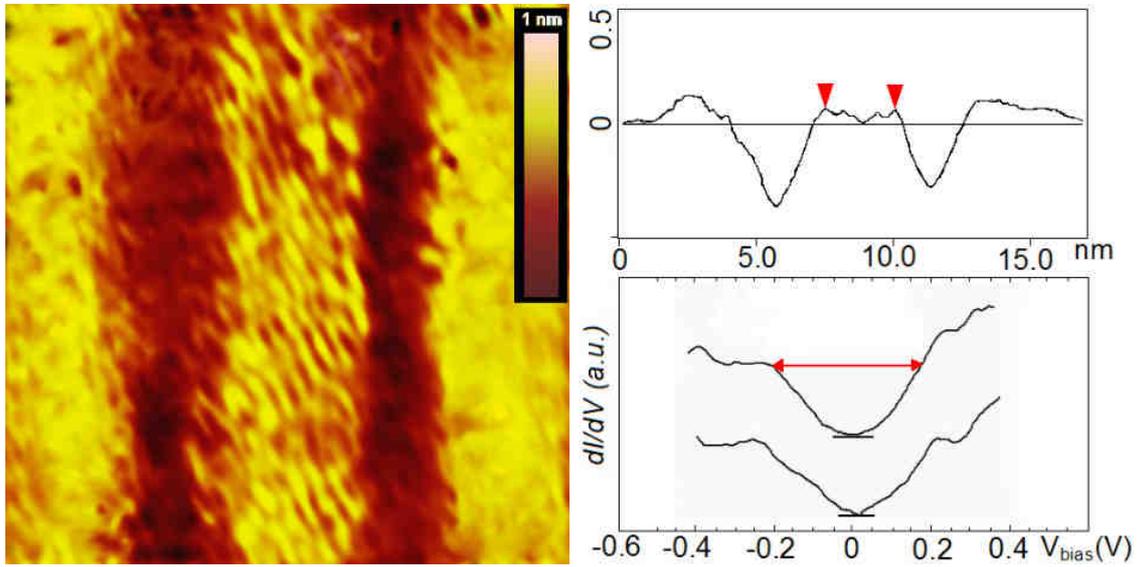

**Figure 5** GNRs for room temperature electronics. **a)** STM image (15x15 nm$^2$, 1 nA, 100 mV) of a 2.5 nm wide armchair GNR. **b)** Average line cut of the STM image revealing the real width of the ribbon. **c)** Representative STS spectra taken on the narrow ribbon showing an energy gap of about 0.5 eV (zero DOS marked by horizontal lines).